\documentclass[12pt]{article}
\usepackage{graphicx}
\usepackage{subfigure}
\usepackage{amsmath}
\usepackage{amssymb}
\usepackage{float}
\restylefloat{table}
\usepackage[bottom=0.5in]{geometry}
\usepackage[symbol]{footmisc}

\begin{document}

\begin{center}
\begin{large}
\title\\{ \textbf{Improved wave function for heavy-light mesons in QCD potential model approach and parameterization of the Cornell potential.}}\\\

\end{large}

\author\

\textbf{$Abdul\;Aziz^{\emph{1,2}},\;Sabyasachi\;Roy^{\emph{2}}\footnotemark\:\;and \;Atri\;Deshamukhya^{\emph{1}}$} \\\
\footnotetext{Corresponding author. e-mail :  \emph{sroy.phys@gmail.com}}
\textbf{1}. Department of Physics, Assam University, Silchar, Assam-788011, India.\\
\textbf{2}. Centre for Theoretical Physics, Department of Physics, Karimganj College, Karimganj, Assam-788710, India.\\
\begin{abstract}
We report improved wave function for mesons in QCD potential model approach using multiplicative method for solution of Schr\"{o}dinger equation for the extreme cases of inter-quark separations ($r\rightarrow0$ and $r\rightarrow\infty$). Using the wave function we find the range of the scale factor c of the Cornell potential with reference to the mass scale of B and D sectors heavy-flavoured mesons. With our computed range of c, we also explore the critical inter-quark separation which should govern the choice of parent-child terms in perturbation method out of the linear and Coulombic terms of the Cornell potential. \\
\end{abstract}
\end{center}

\section{Introduction}\rm
The potential model approach has been very successful and reliable in the studies of the properties of mesons, specially, heavy sector mesons \cite{Eichten}. It is well known that the Cornell potential is the well accepted phenomenological potential to represent quark-antiquark interaction in a meson, which is of the form $V(r)=-a/r+br+c$ \cite{Eichten,Cornell}. With its linear plus Coulombic combination, both the confinement and asymptotic freedom of QCD are preserved within it. With Cornell potential, the static and dynamic properties of mesons are extensively studied by our group recently \cite{SR1,SR2,SR3,SR4,SRmass}. The steps are, first to develop meson wave function using non-relativistic Schr\"{o}dinger equation\cite{Schrodinger} and then by using that equation study the properties of mesons.\\
In the development of meson wave function in such potential model approach using perturbation technique, the choice of parent-child is most crucial. In our earlier work\cite{SR3}, we have taken linear term as parent in perturbation method which resulted in Airy's polynomial function in the wave function\cite{Airy}. In the present work we have improved formalism by developing multiplicative form of the wave function considering the two extreme cases ($r\rightarrow 0$ and $r\rightarrow\infty$)\cite{SR4}.\\
Further, in earlier works, in developing the meson wave function, the scale factor 'c' of the Cornell potential has been taken to be either zero \cite{SR4} or 1 GeV \cite{SR1,SR3} to make it compatible with meson masses in the calculations. However, this seems to be a crude parameterization as the significance of 'c' lies in scaling the potential to make it suitable for representing meson interaction. With this consideration, in this work we have condidered 'c' to be non-zero, deduce the wave function and find the closest range of 'c' which best suits the calculations for masses of heavy-flavoured mesons \cite{PDG}.\\
Furthermore, there should be valid argument for the choice of parent-child in perturbation technique. However, in works referred above, the condition for the consideration of linear confinement term of potential as parent has not been discussed clearly. It is generally accepted that the Coulombic term is dominant at short distance (ultraviolet energy scale) whereas the linear confinement term controls the potential at large distance (infrared energy scale). Here, the boundary condition is set by $V(r_o)=0$, where $r_0$ is termed as the critical distance. With $\langle r \rangle < r_0$, the Coulombic term can be chosen as parent and with $\langle r \rangle > r_0$, the linear term \cite{Atchison}. $r_0$ being crucial for the choice of parent-child, the calculation of $r_0$ for a particular meson is governed by the choices for parameters involved in the potential ($\alpha_s$, b, c).\\
In this work, using improved multiplicative form of wave function, we compute the range of $r_0$ corresponding to our calculated range of 'c'. This sets the bound for taking linear term or Coulombic term as parent term or perturbation term.\\
We also comment on the restrictions on free parameterizations of coupling constant and confinement parameter appearing in the Cornell potential \cite{KKP1,KKP2}.\\
In $\S$ 2 we discuss the formalism, $\S$ 3 contains calculations and results while summary and concluding remarks are stated in $\S$ 4.
\section{Formalism}
\subsection{Wave function}
The quark-antiquark Cornell potential is of the form:
\begin{equation}
V (r) = - \frac{4 \alpha_s}{3 r} + br + c  \\
\end{equation}
where $\alpha_s$ is the QCD coupling constant, $b$ is the confinement parameter and $c$ is the scale factor. We take $\frac{4 \alpha_s}{3 }=a$ to get the simpler form:
\begin{equation}
V (r) = - \frac{a}{r} + br + c
\end{equation}
The Hamiltonian is:
\begin{equation}
H = -\frac{\nabla^2}{2\mu}- \frac{a}{r}+br+c
\end{equation}
Here $\mu=\frac{m_1m_2}{m_1+m_2}$ is the reduced mass of the meson, $m_1$ and $m_2$ being the masses of the constituent quarks.
The two-body radial Schr\"{o}dinger equation in terms of radial wave function $R(r)$ is:
\begin{equation}
[-\frac{1}{2\mu}(\frac{d^2}{dr^2}+\frac{2}{r}\frac{d}{dr}-\frac{l(l+1)}{r^2})- \frac{a}{r}+br+c]R(r)=ER(r)
\end{equation}
For ground state ($l=0$):
\begin{equation}
[\frac{d^2}{dr^2}+\frac{2}{r}\frac{d}{dr}+2\mu (E+ \frac{a}{r}-br-c)]R(r)=0
\end{equation}
Introducing, $U(r)=r R(r)$, the equation (5) transforms to:
\begin{equation}
\frac{d^2 U(r)}{dr^2}= 2\mu (br -\frac{a}{r}-E^\prime)U(r)\\
\end{equation}
where $E^\prime=E-c$. \\
Now, to extract the wave function, we consider two extreme conditions. \\
\textbf{Case-I ($ r \rightarrow \infty $)} \\
Taking $ r \rightarrow \infty $ , when $1/r$ term vanishes and equation (6) reduces to:
\begin{equation}
\frac{d^2 U(r)}{dr^2}= 2\mu (br-E^\prime)U(r)
\end{equation}
Solution of this equation comes out in terms of Airy's function\cite{SR1,Airy} as:
\begin{equation}
U(r) \sim Ai[\varrho] = Ai[\varrho_1 r +\varrho_0]
\end{equation}
Here, $\varrho= \varrho_1 r + \varrho_0$, with    $ \varrho_1 = (2 \mu b )^{1/3}$ and $\varrho_0 = - (\frac{2\mu}{b^2})^{1/3} E^\prime $. $\varrho_0$ is the zero of the Airy's function ($Ai[\varrho_0]=0$).\\
\textbf{Case-II ($ r \rightarrow 0 $)} \\
We take $ r \rightarrow 0 $ when $1/r$ term in equation (6) will prevail:
\begin{equation}
\frac{d^2 U(r)}{dr^2}= 2\mu (-\frac{a}{r}-E^\prime)U(r)
\end{equation}
The solution comes out as:
\begin{equation}
U(r) \sim e^{-r/a_0}
\end{equation}
Here, $a_0 = \frac{1}{\mu a}=\frac{3 }{4\mu \alpha_s}$.
We construct the ground state wave function as the multiplication of the solutions of these two extreme cases (equations 8 and 10):
\begin{equation}
U(r) \sim Ai[\varrho_1 r +\varrho_0] e^{-r/a_0}
\end{equation}
With $N$ as the normalisation factor, our radial wave function has thus the form :
\begin{equation}
\Psi(r) = \frac{N}{r} Ai[\varrho_1 r +\varrho_0] e^{-r/a_0}
\end{equation}
Here, it is worthwhile to mention that although Schr\"{o}dinger equation is non-relativistic in nature, it stands for the studies of heavy-light mesons on the ground of large quark masses involved allowing to consider the velocities of heavy particles to be non-relativistic. However, the relativistic effect in the wave function can be incorporated following Dirac prescription\cite{Dirac}, which yields:
\begin{equation}
\Psi_{rel} (r) = \frac{N}{r} Ai[\varrho_1 r +\varrho_0] e^{-r/a_0}(\frac{r}{a_0})^{-\epsilon }
\end{equation}
Here,
\begin{equation}
\epsilon = 1-\sqrt{1-(\frac{4\alpha_s}{3})^2}=1-\sqrt{1-a^2}
\end{equation}
Truncating Airy's infinite series up to $O(r^3)$, the wave function can be simplified to the form [Appendix-A]:
\begin{equation}
\Psi_{rel} (r) = \frac{N}{a_0^{-\epsilon}}[ k_0 r^{-1-\epsilon} + k_1 r^{-\epsilon} + k_2 r^{1-\epsilon} +k_3 r^{2-\epsilon} ]e^{-r/a_0}
\end{equation}
\subsection{Meson mass}
As in our previous works, here also we use the following relation for calculation of pseudoscalar meson mass.
\begin{equation}
M_P = m_Q +m_{\overline{Q}}+ \triangle E  \\
\end{equation}
Where $\triangle E = <H>$. The Hamiltonian operator $H$ is:
\begin{equation}
H=-\frac{\nabla^2}{2\mu}+V(r)
\end{equation}
So, for ground state meson, we have:
\begin{equation}
 \nabla^{2}\equiv \frac{d^{2}}{dr^{2}}+\frac{2}{r}\frac{d}{dr}
\end{equation}
This gives:
\begin{eqnarray}
<H> = <-\frac{\nabla^2}{2\mu}> + <-\frac{4 \alpha_s}{3r}> + <\sigma r > + <c> \nonumber \\
= <H_1> + <H_2> + <H_3> + <H_4>
\end{eqnarray}
The expressions for the specific terms are given below.
Here,
\begin{eqnarray}
<H_1> = \int_0^{\infty} 4 \pi r^2 \Psi_{rel}(r)[-\frac{\nabla^2}{2\mu}]\Psi_{rel}(r)dr \\
= \int_0^{\infty} 4 \pi r^2 \Psi_{rel}(r)[-\frac{1}{2\mu}(\frac{d^2 \Psi(r)}{dr^2}+ \frac{2}{r}\frac{d \Psi_{rel}(r)}{dr})]dr \\
<H_2> = \int_0^{\infty} 4 \pi r^2 (-\frac{B}{r})|\Psi_{rel}(r)|^2 dr = - 4\pi B \int_0^{\infty} r |\Psi_{rel}(r)|^2 dr  \\
<H_3> =  \int_0^{\infty} 4 \pi r^2 (b r)|\Psi_{rel}(r)|^2 dr = 4\pi b \int_0^{\infty} r^3 |\Psi_{rel}(r)|^2 dr  \\
<H_4> =  \int_0^{\infty} 4 \pi r^2 (c)|\Psi_{rel}(r)|^2 dr = 4\pi c \int_0^{\infty}r^2 |\Psi_{rel}(r)|^2 dr
\end{eqnarray}

\section{Calculation and results}
\subsection{Calculation with fixed values of $\alpha_s$:}
Using the wave function developed in this formalism and equations (16) and (19) the  masses for B and D sector heavy-light mesons are calculated taking the strong coupling constant $\alpha_s$ as 0.39 and 0.22 at charm and bottom mass scale respectively, the value of the confinement parameter b is taken to be 0.183 $GeV^2$ \cite{Eichten}. The input quark masses are taken from ref \cite{Vinodkumar}, the  masses for pseudoscalar B and D mesons are taken from PDG\cite{PDG}.\\
It is worthwhile to mention that the scale factor c is there in the wave function being within the function $\varrho_0$ and also in $<H_4>$. In this calculation non-zero values of c are taken. The meson masses are computed for different values of the scale factor c within the range  $-1.5<c<1.5$. The results for $D_0$, $D_s$, $B_0$ and $B_s$ mesons are shown in Figure-1.\\
\begin{figure}[H]
    \centering
    \includegraphics[width=5 in]{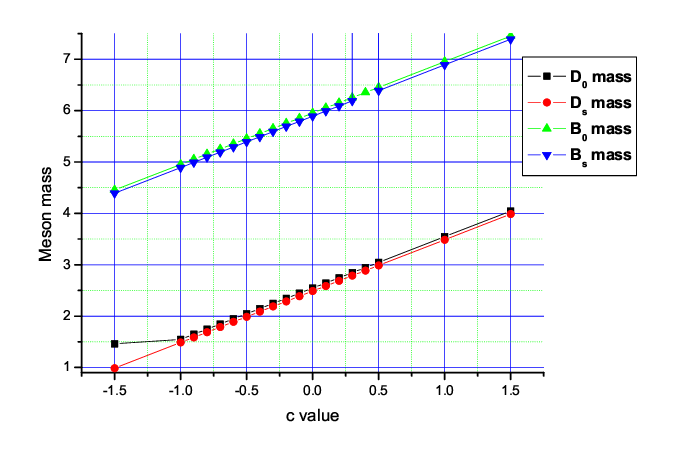}
    \caption{Variation of meson masses with c.}
\end{figure}
The figure indicates that the masses of meson increases almost linearly with the increasing value of 'c'. We then fixed the meson masses at the PDG values and computed the fitting valuees of 'c'. The result is shown in Table-1 and Figure-2. \\
\begin{table}[H]
\begin{center}
\caption{Value of parameter c fitting to PDG masses.}
\begin{tabular}{|c||c||c|}
  \hline
    Meson & PDG mass (GeV) &  c value (GeV) \\
   \hline
$D^0 (c\overline{u})$         & 1.8648 &	-0.6812  \\
$D_s (c\overline{s})$         & 1.9683 &	-0.5188  \\
$B^0 (d\overline{b})$         & 5.2797 &	-0.6746  \\
$B_s (s\overline{b})$         & 5.3669 &	-0.5251  \\

  \hline
\end{tabular}
\end{center}
\end{table}
\begin{figure}[H]
    \centering
    \includegraphics[width=5 in]{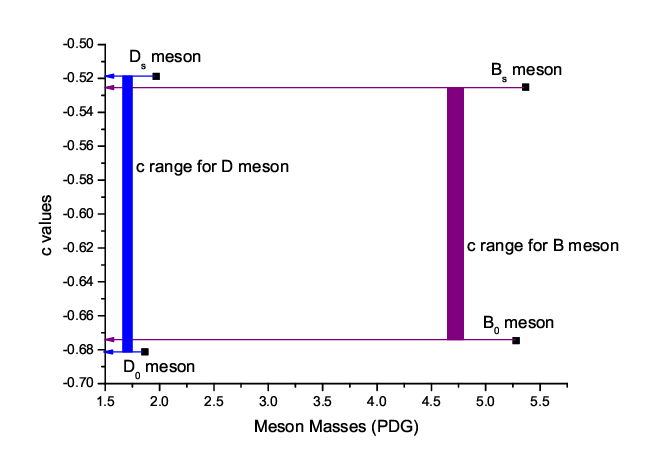}
    \caption{Values of c fitted to standard meson masses.}
\end{figure}
From Table-1, we find that the scale parameter 'c' in the Cornell potential for mesons best suits in the range $-0.5 \;GeV$ to $-0.7 \;GeV$ for the standard PDG masses of B and D sector heavy-light mesons. Whereas for mesons with s-quark ($D_s$, $B_s$) the c value is near $-0.7$ $GeV$, for lighter variants ($D^0$, $B^0$) it is near $-0.5$ $GeV$. This remains within our initial choice for the range $-1.5<c<1.5$. This computed range of c does not also overshoot the range considered in \cite{KKP1}, but refines the range with valid justification.\\
Although the formalism following which the wave function is developed here does not involve perturbation method and hence the choice of parent-child does not arise, still we proceed to study the validity condition for the choice of parent-child out of linear and Coulombic terms of the potential in perturbation method with that computed range of c. Such a condition basically should emanate from the condition of zero potential, i.e, $V(r_0)=0$, where $r_0$ is the the critical distance, i.e,  threshold inter-quark separation of the meson which sets the limit for the choice of linear parent or Coulombic parent. It is generally accepted that\cite{Atchison} if $r<r_0$ Coulombic part of the potential should be taken as parent whereas for $r>r_o$, the linear confinement term should be the parent term in perturbation technique.\\
The variation of $r_0$ with scale factor c is shown in Table-2. \\
\begin{table}[H]
\begin{center}
\caption{Variation of $r_0$ with c (with constant values of $\alpha_s$ and b).}
\begin{tabular}{|c||c|}
  \hline
    c &   $r_0 (GeV^{-1}$) \\
   \hline
-1.0   & 5.943  \\
-0.7   & 4.462\\
-0.6   & 3.990\\
-0.5   & 3.536\\
0.0    & 1.686\\
0.5    & 0.804\\
1.0    & 0.478\\
\hline
\end{tabular}
\end{center}
\end{table}

From the table, it is observed that for c in the range $-0.7< c < -0.5$, the $r_0$ range comes out as $4.5>r_0>3.5$. Given the hadron size, if we take the lower limit of the range to be the critical distance, then with linear parent, $r$ should be greater than $3.5 GeV^{-1}$ and with Coulombic parent it must be smaller than this value.\\
\subsection{Calculation with allowed range of $\alpha_s$:}
In the above calculation we have worked with constant values of $\alpha_s$ for b and c flavoured mesons (0.22 for b flavoured and 0.39 for c flavoured)\cite{Eichten,Alphas, Allen}. We now explored the range of c further by working with the allowed range of $\alpha_s$ rather than a fixed value. We work with ranges of $\alpha_s$ to be 0.19 to 4.2 \cite{Deur, Allen} and our above found range of c ($-0.7$ to $-0.5$), keeping b value fixed at 0.183 $GeV^{-1}$. The result is shown graphically in Fig-2.\\
\begin{figure}[H]
    \centering
    \includegraphics[width=3 in]{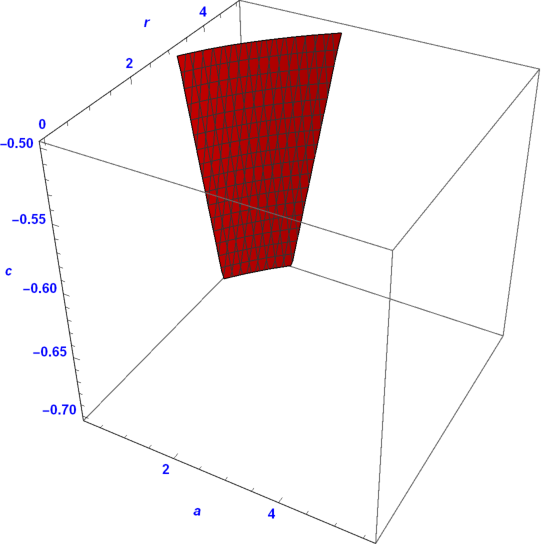}
    \caption{The range of $r_0$ for the given ranges of $\alpha_s$ and c (Here, $a=\frac{4\alpha_s}{3}$).}
\end{figure}
From the graph, the range of $r_0$ is obtained to be 3.17 to 4.5. Thus, we infer that with the variation of $\alpha_s$ within the allowed range, the lower bound of the critical inter-quark separation comes down to to 3.17 from 3.5 as found in our initial calculation with fixed value of $\alpha_s$ in this present work.
\section{Conclusion and remarks}
In the work, we have mainly developed improved meson wave function with the complete form of the Cornell potential considering non-zero value of scale factor c. With that we have explored the range of c that fits in the calculation of the masses of the B and D sector mesons. \\
\begin{enumerate}
  \item We find that c has reasonable effect on the masses of the mesons and as such is also a deciding term of the Cornell potential in the studies of meson properties. It serves as a fitting parameter in computing static and dynamic properties of mesons.
  \item Our computed range of c comes out to be improvement of the works\cite{KKP1,KKP2} in which the c range has been taken on random trial basis. Also, our range of c is in conformity with the value used in work on flux tube model for hybrid charmonium states \cite{Willard}.
  \item Although in this work we have avoided perturbation method in finding wave function for mesons, but to make the picture of the choice of parent-child clear in that method, we have studied the range of critical inter-quark separation $r_0$ which sets the boundary for that choice. We find that taking $\alpha_s$ within 0.19 and 4.2\cite{Deur} and our computed range of c with b to be 0.183 $GeV^{-1}$ the lower bound of $r_0$ comes down to 3.17 from 3.5 which we have obtained working with constant values of $\alpha_s$ for B and D mesons.
  \item It is to be mentioned that in recent work on parameterisation of Cornell potential \cite{KKP2} all of c, b and $\alpha_s$ parameters have been kept as free parameters treating these to be fitting parameters. However, the choice of $\alpha_s$ and b values have some guiding references on which we relied upon in our present work. For $\alpha_s$ we have taken the range considering its governing factors like size of the hadron and momentum transfer \cite{Deur,Alphas,Allen,Eichten}. Regarding the confinement parameter, we would like to mention that it resembles the string tension of string potential for mesons which controls the slope and intercept of the Regge trajectory \cite{Regge}. There is as such not much flexibility for the choice of its value. In our work, we have chosen to keep it fixed at 0.183 $GeV^{-1})$ as predicted in holographic QCD which fits remarkably well with quarkonium data \cite{Deur}.
  \item For the scale factor c in the potential, we have started with a possible range of $-1.5$ GeV to $1.5$ GeV and by fitting our computed meson masses with PDG masses\cite{PDG} we have computed a refined range of c.
\end{enumerate}
Finally, it is worth commenting that the bounds of the critical inter-quark separation $r_0$ inspires us to compare the results of meson properties for both the choices of linear parent and Coulombic parent in perturbation technique. This will clear the ambiguity prevailing in the choice of parent-child. This work is presently in progress.

\appendix

\numberwithin{equation}{section}

\begin{center}
\textbf{Appendix-A}
\end{center}
Airy's infinite series as a function of $\varrho =\varrho_1 r +\varrho_0 $ can be expressed as \cite{Airy}:
\begin{eqnarray}
Ai[\varrho_1 r + \varrho_0] = a_1[1+\frac{(\varrho_1 r + \varrho_0)^3}{6}+\frac{(\varrho_1 r + \varrho_0)^6}{180}+\frac{(\varrho_1 r + \varrho_0)^9}{12960}+...]- \nonumber \\
 b_1[(\varrho_1 r + \varrho_0) +\frac{(\varrho_1 r + \varrho_0)^4}{12}+\frac{(\varrho_1 r + \varrho_0)^7}{504}+\frac{(\varrho_1 r + \varrho_0)10}{45360}+...]
\end{eqnarray}
\begin{flushright}
 with $a_1=\frac{1}{3^{2/3} \Gamma(2/3)}=0.3550281$ and $ b_1=\frac{1}{3^{1/3} \Gamma(1/3)} =0.2588194.$ \\
\end{flushright}
We consider Airy's series up to $O(r^3)$:
\begin{equation}
Ai[\varrho] = a_1[1+\frac{(\varrho)^3}{6}]- b_1\varrho
\end{equation}
From this, we get the Airy function as an explicit function of $r$ as:
\begin{equation}
Ai[r] = k_0 + k_1 r + k_2 r^2 +k_3 r^3
\end{equation}
with $k_i s$ having their explicit form as given below:
\begin{eqnarray}
k_0 = a_1 + \frac{a_1 \varrho_0 ^3 }{6} - b_1 \varrho_0 \\
k_1 = \frac{a_1 \varrho_0^2 \varrho_1 }{2} -b_1 \varrho_1 \\
k_2 = \frac{a_1 \varrho_0 \varrho_1^2}{2} \\
k_3 = \frac{a_1 \varrho_1^3 }{6}
\end{eqnarray}
With this truncated expression of Airy's function, we now construct the wave function as:
\begin{eqnarray}
\Psi_{rel} (r) = \frac{N}{r} [k_0 + k_1 r + k_2 r^2 +k_3 r^3] e^{-r/a_0}(\frac{r}{a_0})^{-\epsilon }\\
= \frac{N}{a_0^{-\epsilon}}[ k_0 r^{-1-\epsilon} + k_1 r^{-\epsilon} + k_2 r^{1-\epsilon} +k_3 r^{2-\epsilon} ]e^{-r/a_0}
\end{eqnarray}
It is to be mentioned that the wave function with Airy's polynomial series brings in divergences in carrying out the integrations with infinite
upper limit in the calculation of  $<H>$. For that we have used truncated Airy series. Regarding sensitivity of the order of polynomial approximation of the Airy's infinite series, it is to be mentioned that Airy's function and its derivatives fall very sharply and almost dies out with increasing r-value ($Ai[3]=0.0066$)\cite{SR1}. This allows us to truncate the infinite Airy series up to third order of $r$ without compromising the result.

\end{document}